# AI Support Meets AR Visualization for Alice and Bob: Personalized Learning Based on Individual ChatGPT Feedback in an AR Quantum Cryptography Experiment for Physics Lab Courses

Atakan Coban[1], David Dzsotjan[1], Stefan Küchemann,[1] Jürgen Durst[2],
Jochen Kuhn[1] & Christoph Hoyer[1]

[1]Chair of Physics Education, Faculty of Physics, Ludwig-Maximilians-Universität München (LMU)
[2]Physics Laboratory Courses, Faculty of Physics, Ludwig-Maximilians-Universität München (LMU)

email: atakan.coban@lmu.de

**Abstract**

Quantum cryptography is a central topic in the quantum technology field that is particularly important for secure communication. The training of qualified experts in this field is necessary for continuous development. However, the abstract and complex nature of quantum physics makes the topic difficult to understand. Augmented reality (AR) allows otherwise invisible abstract concepts to be visualized and enables interactive learning, offering significant potential for improving quantum physics education in university lab courses. In addition, personalized feedback on challenging concepts can facilitate learning, and large language models (LLMs) like ChatGPT can effectively deliver such feedback. This study combines these two aspects and explores the impact of an AR-based quantum cryptography experiment with integrated ChatGPT-based feedback on university students' learning outcomes and cognitive processes. The study involved 38 students in a physics laboratory course at a German university and used four open-ended questions to measure learning outcomes and gaze data as a learning process assessment. Statistical analysis was used to compare scores between feedback and non-feedback questions, and the effect of ChatGPT feedback on eye-tracking data was examined. The results show that ChatGPT feedback significantly improved learning outcomes and affected gaze data. While the feedback on conceptual questions tended to direct attention to the visualizations of the underlying model, the feedback on questions about experimental procedures increased visual attention to the real experimental materials. Overall, the results show that AI-based feedback draws visual attention towards task-relevant factors and increases learning performance in general.

## 1    Introduction

Quantum technology, which is based on the fundamental principles of quantum mechanics, is driving revolutionary advancements in sectors from computing to telecommunications. Quantum computing technologies utilize the principles of quantum physics to address complex problems ranging from cryptography to optimization and quantum simulation by performing calculations at speeds that are unattainable by classical computers [1]. In particular, integrated quantum photonics improves the generation and control of quantum states at the chip scale, setting new boundaries for quantum communication and information processing applications [2]. This technology not only improves the manipulation and stability of quantum states but facilitates more efficient and secure communication networks [3].

Quantum cryptography, which is integral to secure communication, is a very important and rapidly developing quantum technology [4]. It is based on fundamental concepts from quantum physics, such as its probabilistic nature, the polarization state of a single photon, and the principle of superposition. Therefore, integrating quantum cryptography effectively into real-life applications requires a high level of quantum physics knowledge and application skills. Intensive, effective quantum physics education is the primary way to achieve this.

The integration of quantum technologies, such as quantum cryptography, into existing telecommunications infrastructure presents various obstacles. One of the most significant current challenges is the shortage of human resources who can effectively manage quantum communication infrastructures, which is defined in the literature as the 'quantum skills gap' [5]. One of the main reasons for this gap is that quantum physics education has fallen behind technological advancements. Therefore, the primary and most crucial method to address this gap is to include comprehensive quantum physics training in education [6, 7].

One approach to imparting knowledge about quantum cryptography is a physics lab course. As quantum cryptography is an extremely abstract and complicated topic, learners could be provided with virtual visualizations of quantum states via AR within the in-person setup. In this paper, we adopt this approach and examine how much individualized feedback provided by ChatGPT can improve the answers of participants interacting with an AR quantum cryptography experiment. Eye tracking data analysis provided additional insights into the learners' 3D visual attention during the lab course and the effects of verbal ChatGPT feedback on their gaze.

The integration of ChatGPT-based feedback into an AR-based quantum physics experiment and the analysis of the effects of that feedback in terms of learning outcome and learning process using gaze data are the key elements that make this study significant.

## 2 Theoretical background

### 2.1 Quantum cryptography: A brief overview

Methods for encrypting and sending sensitive information between two parties so that a third party cannot access the content have been pursued in various human cultures and societies since the dawn of human history. Practically, all of these methods or protocols have two key points in common: 1) a key or rule is used to encrypt the message on the transmitting side and decrypt it on the receiving side, and 2) both sides must possess the key, but no one else can have it. The security of the protocol depends on the strength of the encryption and the likelihood that no one other than the sending and receiving parties can acquire the encryption key.

Next, we will describe the BB84 protocol, a key distribution method that relies on the laws of quantum mechanics and is, in principle, perfectly secure.

#### 2.1.1 One-time pad encryption

At the heart of the BB84 protocol lies a classical method for transmitting an encrypted message: the one-time pad encryption, or single-use key encryption, method. The principle can be demonstrated with a simple example. Suppose we want to encrypt a message that is the letter A. We add its binary ASCII encoding and a randomly generated binary key (mod1 addition) to obtain the encrypted message:

$$message\ (A): 01000001, key: 01010110 \qquad (1)$$

$$encrypted\ message: 01000001 + 01010110 = 00010111,$$

which can then be decoded on the receiving side by adding the encryption key to it again:

$$recovered\ message: 00010111 + 01010110 = 01000001 \qquad (2)$$

resulting in the message that we wanted to encrypt. The encryption key must be used one time only, i.e., for each message string, a new random key must be generated. This classical method, in principle, is 100% secure, provided that only the sending and the receiving party have the encryption key. The problem of secure data transmission is therefore reduced to transmitting an encryption key from the sender to the receiver such that no third party can intercept it. The use of quantum bits within the BB84 protocol and the laws of quantum mechanics make this possible and even permit the presence of a potential eavesdropper to be detected.

*2.1.2  Key distribution and BB84 protocol*

In the BB84 protocol, we surpass mere 1s and 0s and introduce the concept of bases. This is conveniently demonstrated in our experiment, where polarized photons serve as quantum bits (qubits). With lambda/2 plates, the polarization of incoming linearly polarized light can be rotated at will to generate an outgoing photon of arbitrary linear polarization. Consider two sets of photonic polarization basis states:

$$\{|H>, |V>\}, \{|+45>, |-45>\}, \qquad (3)$$

where the first pair are the horizontal and vertical polarization states and the second are the states of polarization in the +45- and -45-degree directions, respectively. We name the bases + and x, respectively, and define two sets of qubits as

$$|0>_+ = |H>,\ |1>_+ = |V>,$$

$$|0>_X = |-45> = 1/\sqrt{2}(|H> - |V>), \qquad (4)$$

$$|1>_X = |+45> = 1/\sqrt{2}(|H> + |V>)$$

Thus, the transmitted information has two layers: the value of the bit (0 or 1) and the value of the base (+ or x).

The quantum key distribution within the BB84 protocol happens as follows. The sending party (Alice), by rotating their lambda/2 plate, generates photons with random bases (+ or x) and bit values (0 or 1). The receiving party (Bob), in turn, rotates their own lambda/2 plate, randomly choosing whether to measure in the + or the x basis, and measures an incoming string of 0s and 1s. For instances where both Alice and Bob set their bases to the same value, then a 1 sent by Alice will always be measured as a 1 by Bob. However, if they set different bases, say, Alice sends $|1>_X$ and Bob's measuring device is set in the + base, then, according to Eq. (4), this incoming photon on Bob's side will be randomly mapped either to the 0 or the 1 of the + base. Crucially, as single photons are being sent, Bob's detectors never click simultaneously: He either measures a 0 or a 1. The correspondents can be sure that Bob's measurements were correct when his choice of base coincides with Alice's choice of base, e.g., a 1 sent by Alice was measured as a 1 by Bob. Eventually, both Alice and Bob will have a list of base values and bit values. As the last step of the quantum key distribution, they use a classical information channel to compare their base sequences but not their bit values. The qubit transmissions where their bases did not match are discarded and the remaining string of 0s and 1s serve as the shared key ready to be used in the encryption.

*2.1.3    Detection of an eavesdropper*

The detection of the presence of an eavesdropper is where quantum key distribution truly shines. Suppose that a third person, Eve, can capture the stream of photons sent by Alice. As the no-cloning theorem states, she cannot read out the photon sent by Alice and simultaneously send an identical photon to Bob: She must generate new photons which she then passes on to Bob. Therefore, Eve has both a detector and an emitter that work in both + and x bases and can measure and generate 0s and 1s in these respective bases.

With Eve present, even when Alice's and Bob's bases coincide (both either + or x), their bit values will not always be equal! As Table 1 shows, at every instance when Eve's emitter base varies from Alice's and Bob's bases, there is a 50% chance that Bob measures a different bit value than what Alice sent. As the base choices are completely random and Alice's and Bob's bases are equal, in an average of 50% of the cases, Eve's base will not match Alice's and Bob's (see Table 1). In cases of a base mismatch between Eve and the other two, there is a further 50% chance that Bob will measure a different bit value than that sent by Alice. Therefore, assuming Alice's and Bob's bases match, the overall probability of wrong bit transmission is the product of these chances, i.e., 25%.

**Table 1** Probability of matching bit values for Alice and Bob, for several different base choices, provided Alice and Bob choose the same base. Eve's base will not match Alice's and Bob's in about 50% of the cases.

| Alice's and Bob's bases | Eve's emission base | Probability that Alice's and Bob's bit values match |
|---|---|---|
| ++ | + | 100% |
| ++ | X | 50% |
| XX | + | 50% |
| XX | X | 100% |

To detect the potential eavesdropper, after comparing the bases and discarding the irrelevant instances, Alice and Bob also publicly compare some of the remaining bit values. If they see a bit value mismatch in about 25% of the cases, they can reliably state that security has been compromised and discard the quantum key. This happens before any part of the message itself has been transmitted, giving Eve no way of accessing it.

**2.2    Physics education research on quantum cryptography**

Quantum physics concepts cannot be directly observed due to their abstract nature [8, 9]. The mathematical formalism of quantum mechanics adds another layer of complexity to the topic [10]. These factors make it difficult for students to contextualize quantum physics concepts and relate them to real-world phenomena, creating significant barriers to the learning process [11].

In particular, learning quantum cryptography requires cognitive skills and practical application of the related concepts, in addition to the challenges of learning quantum physics, further complicating the learning process [12]. Effective learning in this field entails visualizing abstract concepts to ensure students' active participation in the process and show where these concepts can be used in real life through practical processes [13]. Therefore, supporting quantum cryptography education with visual tools and interactive methods is believed to improve learning outcomes; as a result, hands-on experiments, games and interactive simulations have become very popular for teaching quantum cryptography [14-17].

Vadla et al. (2019) reported that QuaSim [15], a gamified intelligent tutoring system developed for teaching quantum cryptography, had positive effects on students' achievement and interest in processes involving their active participation [18]. Utama et al. (2020) allowed students at the pre-university level to implement the BB84 protocol with the experiment set they developed for the high-school level and received feedback from students [19]. The students expressed satisfaction with the application and appreciated the opportunity to practice the concepts they had learned. Pallotta (2022) stated that the QuVis [17] simulation was greatly effective in supporting students in analysing the different elements of the key distribution process and that students evaluated this type of activity very highly as it helped them understand how quantum physics is not only an abstract or intuitive theory but also a technological resource [20].

An activity focused on understanding how to build a secure protocol with quantum cryptography applications has dual advantages of raising awareness about the potential of quantum technologies as well as creating a learning context to deepen students' understanding of quantum superposition and quantum measurement [20]. Such activities allow students to consolidate their theoretical knowledge in practice and may both raise their awareness of quantum technologies and help them to better grasp the basic principles of quantum mechanics [21].

Therefore, experimental processes involving quantum cryptography are important both for students to have the opportunity to realize quantum applications in the field and learn about quantum concepts by observing how they are used in practice. Thorlabs has developed the *Quantum Cryptography Analogy Demonstration Kit*, an experiment set for the BB84 protocol on quantum cryptography suitable for the high school and university levels [22].

The experimental set allows students to learn how to determine a secret keycode by following the BB84 protocol. Through this collaborative research-based approach, they can gain insights into the probabilistic structure of the quantum world, the disruption of quantum systems by observation, and the idea of data transmission with quantum bits [23].

As the underlying physical concepts are highly complex and difficult for learners to understand, an adapted version of the original setup that provides individual feedback and intuitively understandable real-time representations of the invisible quantum states might be beneficial. In such an adapted version of the experiment, augmented reality glasses would show virtual visualizations of quantum states that would otherwise not be directly observable, making them visible within the real-world setup. Furthermore, an option for verbal interaction with ChatGPT could offer automated and, primarily, individual feedback on students' answers. Such feedback may drive the student group to discuss questions further, improving the accuracy of their answers [24].

The learning method employed in this study combines hands-on experimental activities with virtual visualizations, with ChatGPT providing feedback to support students' learning outcomes. Before discussing the study in greater detail, an overview of the theories related to learning with multiple representations, augmented reality, and formative feedback will be presented.

## 2.3 Multiple representations and augmented reality

For scientific learning, representing content through different types of representations, or multiple external representations (MERs), is well-documented as beneficial in the natural sciences [25] and physics [26]. This holds especially true for abstract content, such as quantum physics. MERs are especially important for conceptual understanding [27] and are discussed as a necessary condition for in-depth understanding [28].

For visualizing MERs, technology-enhanced learning environments have often been used, e.g. simulations (for quantum physics; [29]). AR technology combined with real-world experimental setups has become increasingly prevalent in various fields, e.g. to provide 3D visualizations and enhance the real-world environment with overlaid virtual content [30, 31]. AR systems' key elements include the alignment of virtual 3D representations with physical objects in the real world and an interactive 3D visualization that changes in real time based on interactions with these physical objects [32].

There is broad evidence and empirical consensus that AR environments foster learning in general [33, 34]. A recent second-order meta-analysis compared AR-based learning environments to traditional learning environments without AR [35]. The mean effect size can be interpreted such that all individual analyses demonstrate medium to large effects and are, therefore, consistent regarding the positive impact of AR on learning achievement. AR-driven learning activities simplify students' mental processes [30] and enhance their understanding of physical phenomena [36]. In quantum physics laboratory settings, AR technology can provide a more understandable conceptual process by presenting abstract quantum physics concepts that cannot be observed directly as virtual models and allowing users to interact with them [37]. Moreover, AR departs from two-dimensional displays by allowing concepts to be observed in three dimensions and interacted with. This shift to spatial and interactive learning can have profound effects on conceptual understanding [38]. Such environments can help students engage with complex quantum phenomena more tangibly and intuitively, enhancing their understanding and involvement in the subject [39]. Therefore, teaching processes that involve visualizing unobservable quantum physics concepts with three-dimensional virtual representations connected to the real world via AR and applications in which students can interact with these virtual representations and observe their changes instantaneously can have an increasing effect on student success. See Dzsotjan et al. (2024) for a detailed discussion of the MER used in this AR environment [40].

## 2.4 Formative feedback via large language models

In the field of education, LLMs can be utilized in various beneficial ways, such as delivering lectures, solving examples, providing personalized assistance and developing educational materials [41]. LLMs can provide formative feedback to students [42]. In education, personalized feedback adapted to a student's learning situation is crucial, as it fundamentally supports students in monitoring their progress and making meaningful connections with the material [43]. Feedback in the educational environment improves learning outcomes [44-46] and is of significant importance in quantum physics education. Due to the abstract nature and advanced mathematical structure of quantum physics, identifying and addressing students' confusions and learning difficulties during the learning process is essential, as is asking students questions throughout the process and providing feedback based on their responses to diagnose their current understanding and enhance their learning processes [9]. This is vital for preventing and correcting learning difficulties that may arise from incorrect understandings. This approach creates a more structured and effective educational experience, facilitating deeper and more sustained learning outcomes.

Bouchée et al. (2023) highlighted the need for students learning quantum physics to receive high-quality feedback from their teachers [9]. This feedback makes students aware of misconceptions in their learning, guiding them towards more effective strategies for understanding quantum physics. In a related study, Abdurrahman et al. (2018) investigated the impact of feedback in teaching quantum physics [47]. Their findings indicated that feedback in quantum physics education significantly improved the scores of physics teacher candidates on the quantum physics concept inventory test. These studies emphasize the critical role of effective feedback in enhancing students' understanding and performance in quantum physics, supporting more accurate and deeper learning.

Some of the biggest challenges in giving effective feedback arise from teachers' limited time and heavy workloads or negative teacher or student attitudes [48, 49]. Giving students individual feedback, especially in crowded classroom environments, requires serious time and intense teacher performance, and is often impossible in practice [50, 51].

LLMs can analyse students' responses and leverage their theoretical awareness of effective feedback to provide comprehensive and varied comments [42], and they have potential in flipped classrooms [52]. Steinert et al. (2023) developed a platform that uses the ChatGPT language model to provide formative feedback to students to enhance various self-regulated learning processes [42]. On this platform, students answer questions provided by their teacher and may improve their answers after receiving personalized formative feedback from ChatGPT. Similarly, Uchiyama et al. (2023) developed a platform where ChatGPT gives feedback on students' answers for questions related with an example situation via video; the researchers stated that this platform could eliminate the difficulties of giving feedback in flipped classrooms [52].

Jansen et al. (2024) stated that the feedback given by LLMs was comparable in quality to the feedback given by experts [53]; however, Steiss et al. (2023) emphasized that human feedback was clearly superior compared to the feedback given by GPT 3.5 [54]. Jacobsen and Weber (2023) argued that GPT-4 may surpass teachers in terms of general quality if trained with effective prompts [55]. Dai et al. (2023) found that ChatGPT can produce more detailed feedback that summarizes students' performance more fluently and consistently than human instructors, and, based on this, they stated that ChatGPT can provide feedback that will help students complete tasks and improve their learning skills [56].

Feedback can be given in different formats, such as written or verbal feedback. Agricola et al. (2020) noted that verbal feedback had a significantly higher effect on students' perception of the feedback than written feedback, and a high perception of feedback is important for the feedback to be effective [57]. Merry and Orsmond (2008) stated that students found verbal feedback more positive and that verbal feedback was easier for them to understand because it was closer to dialogue [58]. Van der Schaaf (2013) gave written feedback to a group of 72 students, and then half of that group was given verbal feedback later in the study; they concluded that the students who received verbal and written feedback found it more effective than the students who did not receive verbal feedback [59].

Therefore, LLMs can be effectively integrated into personalized environments where they provide students with formative feedback. This allows them to overcome teachers' difficulty in giving personalized feedback to each student in a classroom environment. In addition, adding LLMs to AR environments and being able to conduct spoken dialogues with them allows them to effectively support students during experiments. Considering these potential benefits, LLMs' ability to be added to the AR environment to communicate verbally with students and give them feedback has a high potential to improve students' learning gains by facilitating personalized learning processes.

## 2.5 Eye-tracking

The eye–mind assumption states that gaze fixations and cognitive processing of the fixated content are closely linked [60, 61]. Accordingly, eye tracking is frequently used in learning sciences to investigate learning processes in detail [62-65]. Eye-tracking methods have already provided exciting insights into learning processes in physics education [66]. For example, during linking processes between the laboratory system and the coordinate system, a continuous animation that visualizes the translation process binds the gaze very strongly, whereas interactive measurement documentation tasks with feedback result in many cross-representational eye movements, which accompanied a more successful learning process [67]. Eye-tracking studies have also helped illuminate learning processes about vector fields [68,69], graphs [70-73] or interactive learning systems in quantum physics [29]. In classical eye tracking on a 2D computer screen, research focused primarily on analysing the total number or duration of gaze fixations. In addition, the gaze duration in 'areas of interest,' i.e. specific sections of the 2D learning content, was often examined. The total number, direction, length and duration of saccadic movements, which are the eye movements connecting one fixation to the next, have also been examined [62, 74].

As this study did not involve a screen-based learning unit, innovative approaches must be chosen to investigate eye movements during real experimentation and interaction with virtual displays. To do so, the ARETT toolkit for the HoloLens 2 (HL2) was developed [75]. It allows volumes of interest to be defined. These volumes are registered as hit when the eye ray of the HL2 collides with the surface delimiting the volume. The toolkit has been used successfully to investigate eye movements in AR learning units and, especially, to analyse learning success [76].

For this study, ARETT was used to determine the total fixation duration on specific elements of the learning environment. In addition to demonstrating the learning effectiveness of AI-based feedback, this provides exciting insights into the gaze behaviours related to the learning processes that are triggered by feedback. Individuals' total fixation durations are associated with various cognitive processes. Studies indicate that total fixation duration may be directly related to tasks' difficulty level. As the difficulty increases, individuals require more time and effort for information processing, which manifests as longer total fixation durations [77]. In addition, when an individual attempts to overcome a challenging task or encounters an unexpected impasse, an increase in total fixation duration is observed [78]. Moreover, an individual's level of experience significantly affects total fixation duration. Experienced individuals, having previously encountered similar tasks or situations, can process information more efficiently, which can result in shorter total fixation durations [79].

Finally, research reveals that total fixation duration is related to the intensity of cognitive processing [80] and, therefore, presents an informative indicator to interpret the effects of AI-based feedback.

## 3 Research questions

This study aims to investigate the effects of feedback provided by ChatGPT in AR-enhanced experiments in physics lab courses. To investigate the effects of integrating ChatGPT as a feedback tool, the following two research questions were analysed in a crossover design:

RQ1: Cognitive learning outcome: Does formative feedback via a large language model lead to a higher conceptual understanding during experimentation?

RQ2: Cognitive learning process: How does adding individualized ChatGPT-based feedback change students' visual attention while answering comprehension questions in a laboratory course on quantum cryptography?

## 4 Methods

### 4.1 Participants

In total, 38 students (age: mean=20.95; SD=2.42), of whom 24 were male and 14 were female, voluntarily participated in the experiment, which lasted about 3 hours.

All participants conducted the quantum cryptography analogy experiment in the laboratory following a lab manual specifically prepared for the experiment. All groups came to the laboratory at different times during the lab course.

### 4.2 Experimental setup

The quantum cryptography analogy experiment set developed by Thorlabs is based on the BB84 protocol. It includes an experimental setup using short-duration laser pulses instead of single photons. The experiment set primarily consists of a laser source, lambda/2 plates and detectors. Students adjust the angles of the lambda/2 plates to determine the bases and bits of the emission and use the detectors to measure the incoming bits.

In the AR-enhanced version of the experiment, vectorial representations are used to visualize the polarization state of a single photon. QR codes are attached to the experimental setup to ensure that the AR glasses display the visualizations in the correct locations of the setup. As soon as the glasses recognize the QR codes, the corresponding visualizations are displayed. A microcontroller reads out the states of the lambda/2 plates in the setup in real time and sends this data to the AR glasses via Bluetooth. This enables the current polarization states in AR to be displayed in real time according to the lab partners' actions.

The AR application incorporates the most immersive features of AR systems, such as combining real and virtual objects in a real environment, operating interactively and in real-time and aligning real and virtual objects with each other [32]. The 3D visuals on the left and right sides of each lambda/2 plate in the real world update dynamically according to the lambda/2 plate's angle. These visualizations allow students to observe the polarization state before and after it reaches the lambda/2 plate, as well as the effect of the angle of the lambda/2 plate on polarization. The AR visualizations included in the application are shown in Figure 1. Elements with a white background show the angles of the lambda/2 plates, and elements with a black background show the photons' polarization state. Both visualizations update according to the angles of the lambda/2 plates.

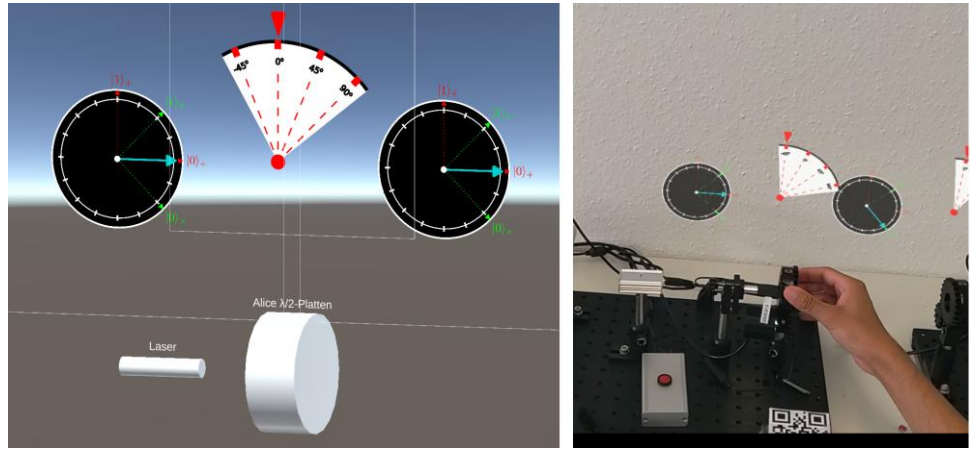

**Figure 1** AR visualization in the application. The elements with a white background show the angle of the lambda/2 plate and update in real time if the physical setup is changed. In the circular elements with a black background, the polarization state of a single photon is represented with vector notation; these also update according to changes in the angle of the real lambda/2 plate.

### 4.3  Objectives in the laboratory course

The primary objective of the quantum cryptography analogy experiment is to exchange bits using the BB84 protocol to generate a key for secure communication and to test, through different methods, whether an agent named Eve was eavesdropping on the exchange. The students followed the protocol twice—once without Eve in the system and once with Eve present. Their task was to determine the key code and test for Eve's presence with two different methods. In addition, at four different times during the experiment, ChatGPT verbally asked the students a question related to the previously completed step.

All steps of the process are detailed in Table 2.

**Table 2** Experimental process steps.

| Step | Task |
| --- | --- |
| 1 | Base-bit determination (for Alice); base determination (for Bob) |
| 2 | Sending and receiving bits (Alice and Bob) |
| 3 | Base comparison, key determination |
| 4 | First comprehension question |
| 5 | Encrypting, transmitting and decrypting a secret message |
| 6 | Second comprehension question |
| 7 | Base-bit determination (For Alice); base determination (for Bob and Eve) |
| 8 | Sending and receiving bits (Alice, Bob and Eve) |
| 9 | Third comprehension question |
| 10 | Base comparison, key determination, selection of a random bit from the key and comparison |
| 11 | Fourth comprehension question |

After listening to the questions in Sections 4, 6, 9 and 11, the students determined their answers and communicated them to ChatGPT. Then, depending on their assigned group, the students either received feedback from ChatGPT or did not (see section 4.5).

## 4.4 ChatGPT usage in the application

During the lab course, students must perform different lab tasks. For specific tasks (see section 4.3) the students interact with ChatGPT by asking questions and receiving answers. To verbally interact with ChatGPT, the speech-to-text feature first converts the students' spoken words into text, which is then sent to ChatGPT as an input. Subsequently, the written output from ChatGPT is converted into speech with a text-to-speech function, allowing students to hear the response. The ChatGPT menu in the application is shown in Figure 2.

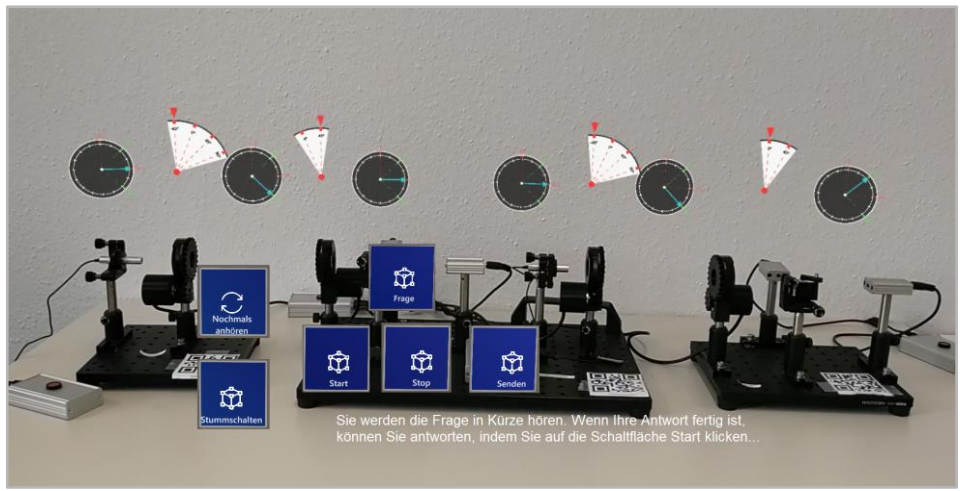

**Figure 2** Interface for interaction with ChatGPT.

To provide feedback, the OpenAI API and the GPT-4 model [81] were used. For ChatGPT to give feedback on students' verbally formulated answers, pre-prompts were sent along with the request.

**i. The prompts prepared for questions with feedback include:**

- The relevant question and its answer.
- Information that the prompt comes from an AR-assisted experiment conducted by students.
- A command to ChatGPT to ask the question without mentioning that it is an AI bot.
- A directive that when ChatGPT receives an answer from students, it should act like a teacher and compare the answer with the correct answer. In cases of incorrect or incomplete responses, it should guide students toward the correct answer without providing information about the correct answer.
- Tell students to proceed to the next stage once their answer is considered correct.

**ii. The prompts prepared for questions without feedback include:**

- Only the relevant question.
- Information that the question comes from an AR-assisted experiment conducted by students.
- A command to ChatGPT to ask the question without mentioning that it is an AI bot.
- A directive to tell students to proceed to the next stage without providing feedback after receiving their response.

### 4.5 Study design

At four points in the lab course, students were asked to answer comprehension questions that referred to previous tasks. The students responded verbally to these questions. The use of ChatGPT for these questions involved two distinct methodologies: with feedback (F) and without feedback (W). In the feedback sections, ChatGPT provided feedback by comparing students' responses with the correct answers. Feedback guides students to improve their answer accuracy through discussion in cases of incomplete or incorrect responses. Students were randomly divided into two groups, A and B. Group A received ChatGPT-based feedback on their answers to Questions 1 and 3 and Group B received ChatGPT-based feedback on their answers to Questions 2 and 4. The research design is shown in Figure 3.

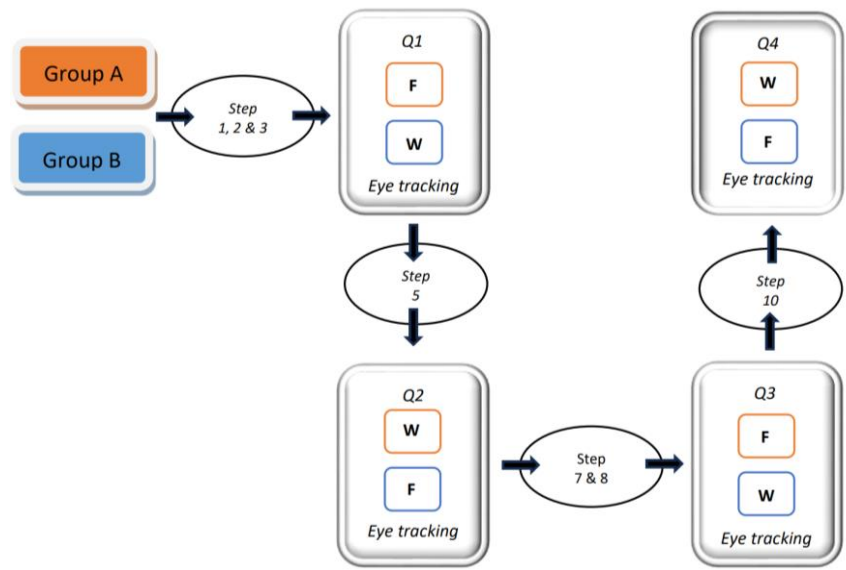

**Figure 3** Research design of the study. Both groups performed the same experimental steps (see Table 2). After completing specific steps, the groups answered Questions 1, 2, 3 and 4 (Q1, Q2, Q3 and Q4), either with feedback (F) or without feedback (W). During the Q/A stages, eye-tracking data were collected from the students.

### 4.6 Instruments

*4.6.1 Cognitive learning outcome: Questions about experimental processes*

The first research question was about the impact of ChatGPT feedback on students' performance scores when answering questions. Each question's answer contained four parts, each part assigned 0.25 points, and a complete explanation of each question was assigned 1 point. These questions were integrated into the prompts sent to ChatGPT, and the students' conversations with ChatGPT were documented in text form. The first question involved explaining why, in quantum cryptography, two bases are used. The second question asked students to name the necessary aspects to generate a key. The third question involved describing the experimental procedure that Eve uses to try to intercept the communication, and the fourth question asked how it is possible to discover whether a spy has intercepted a communication by comparing bits and bases. Overall, questions 1, 2 and 4 refer to more conceptual knowledge, while question 3 refers to an experimental approach.

*4.6.2   Cognitive learning process: Eye gaze data collection with ARETT*

To answer the second research question, eye gaze data were collected with the ARETT toolkit to analyse the students' total fixation duration while responding to ChatGPT feedback. The ARETT toolkit is specially developed to easily collect eye gaze data in AR applications. ARETT consists of a data access layer that makes timestamped gaze data available in real time, a data recorder to store the data, a web-based control interface and a set of auxiliary tools for data visualization. By integrating this toolkit into their applications, developers can collect data in a controlled manner through the web-based interface. Detailed information about ARETT can be found in the study of Kapp et al. (2021) [75].

After adding the ARETT toolkit to the application, specific areas of interest (AOIs) were created for the physical and virtual components of the experiment. AOIs are used in eye-tracking studies to define specific regions where participants are likely to focus their attention. Physical AOIs are connected to the components at experimental set and the virtual AOIs are connected to AR representations.

## 4.7   Data analysis

*4.7.1   Cognitive learning outcome: Data analysis on the effects of ChatGPT feedback*

The file containing written transcripts of students' verbal conversations with ChatGPT throughout the experiment was transferred from HL2 to a computer at the end of the experiment. Necessary aspects that are necessary to answer the question correctly were determined in advance. The scoring was then carried out on the basis of these considerations.

To test for the presence of a carry-over effect, the sum score of the four questions was calculated for both Group A and Group B. Then, these scores were tested for differences between the groups. The carry-over effect is examined to determine whether the different experimental conditions influence each other [82]. In this study, the potential carry-over effect was examined to see whether receiving and not receiving feedback influenced students' work (for example, if the feedback students receive on Q1 affects their answer to Q2). This effect describes the persistence of the impact of one method (receiving feedback on responses) on the subsequent method (not receiving feedback on responses). In crossover design studies, the presence of a carry-over effect can jeopardize the reliability and validity of the results [83]. Therefore, controlling for this effect is essential.

To compare the scores of questions with feedback to those without feedback, the differences (*D*) between the scores for each group were initially calculated with the following equation:

$$D = [(Q1+Q3) - (Q2+Q4)] \qquad (6)$$

Q1 and Q3 are the scores of the questions for which Group A received feedback, while Q2 and Q4 are the scores of the questions for which Group B received feedback. If the scores are high when feedback is received, Group A's *D*-scores will be higher than Group B's *D*-scores. As the difference between the scores with and without feedback increases, the difference between the mean *D*-scores of the two groups will also increase. This allows scores in the feedback and non-feedback conditions to be compared using the *D*-scores. The *D*-scores for Groups A and B were compared with independent sample *t*-test.

*4.7.2 Cognitive learning process: Data analysis for students' total fixation durations during ChatGPT interactions*

The eye-tracking data collected with ARETT was automatically saved within HL2 as a .csv file when the data collection process was completed.

Total fixation duration metrics were used to analyse the collected eye-tracking data. Total fixation duration gives information about visual attention on the predefined AOIs related to physical components and virtual representations while answering the questions. Fixation duration measures the specific time a student spends looking at a particular AOI. Throughout the process, all such gazes are aggregated to calculate the total fixation duration. During students' responses to questions, the total fixation durations were first calculated for each AOI and each participant. Then, these totals for each participant were divided by that participant's total fixation duration for all AOIs to obtain the normalized total fixation duration for specific components of the experiment. The subsequent analyses were conducted using these normalized total fixation durations. This method determined how visual attention was distributed between the two AOIs.

A paired-sample t-test was used for the quantitative analyses related to the gaze data. Since these analyses were conducted only on students who received feedback, the sample size is limited. To minimize the risk of overlooking significant differences, the significance level for the *p*-value was set at 0.10. This choice is based on the requirement for a larger sample size to reliably detect interaction effects, as noted in the study by Cronbach and Snow (1977) [84]. Since the recommended sample size from the power analysis was not fully met, the alpha error level was increased to optimize the beta error and reduce the likelihood of missing interaction effects.

## 5 Results

In this section, the results will be presented. First, the results regarding the effects of ChatGPT feedback on students' answers will be described. Then, the results of the analyses of the normalized total fixation duration will be presented.

### 5.1 Learning outcome results related to the effects of ChatGPT feedback

During the answering process, no limitation was imposed on the amount of feedback students could receive. They could continue receiving feedback on their responses until ChatGPT was convinced that their answer was sufficient. Receiving multiple rounds of feedback indicates that the students' responses were inadequate or incorrect and that they needed more feedback to improve their answers. The number of feedback responses received by the groups is presented in Table 3.

**Table 3** Number of groups related with the amount of feedback received.

|            | Q1 (Group A) | Q2 (Group B) | Q3 (Group A) | Q4 (Group B) |
|------------|:---:|:---:|:---:|:---:|
| **0 feedback** | 6 | 4 | 4 | 7 |
| **1 feedback** | 5 | 2 | 5 | 2 |
| **2 feedback** |   | 3 | 2 | 1 |
| **3 feedback** |   | 1 |   |   |

Table 3 shows that a high percentage of participants responded to Q1 and Q4 without receiving feedback. Most participants received feedback on Q2 and Q3 at least once.

The average scores for each question and the impact of feedback on these scores are presented in Table 4 and the bar graph in Figure 4.

**Table 4** The average score of learning outcome ($\bar{X}$) and standard error (SE) of the points groups received from answering questions. The 'before feedback' values are based on the points students would have received from their initial answers before any feedback was provided. The 'after feedback' values were calculated using the total points from all answers given to the question (for feedback groups).

|    | **Group** | **Before Feedback** | | **After Feedback** | |
| --- | --- | --- | --- | --- | --- |
|    |    | $\bar{X}$ | SE | $\bar{X}$ | SE |
| **Q1** | A | 0.454 | 0.117 | 0.500 | 0.095 |
|    | B | 0.400 | 0.100 | — |    |
| **Q2** | A | 0.523 | 0.132 | — |    |
|    | B | 0.350 | 0.135 | 0.675 | 0.084 |
| **Q3** | A | 0.295 | 0.030 | 0.636 | 0.070 |
|    | B | 0.200 | 0.033 | — |    |
| **Q4** | A | 0.477 | 0.071 | — |    |
|    | B | 0.350 | 0.100 | 0.500 | 0.092 |

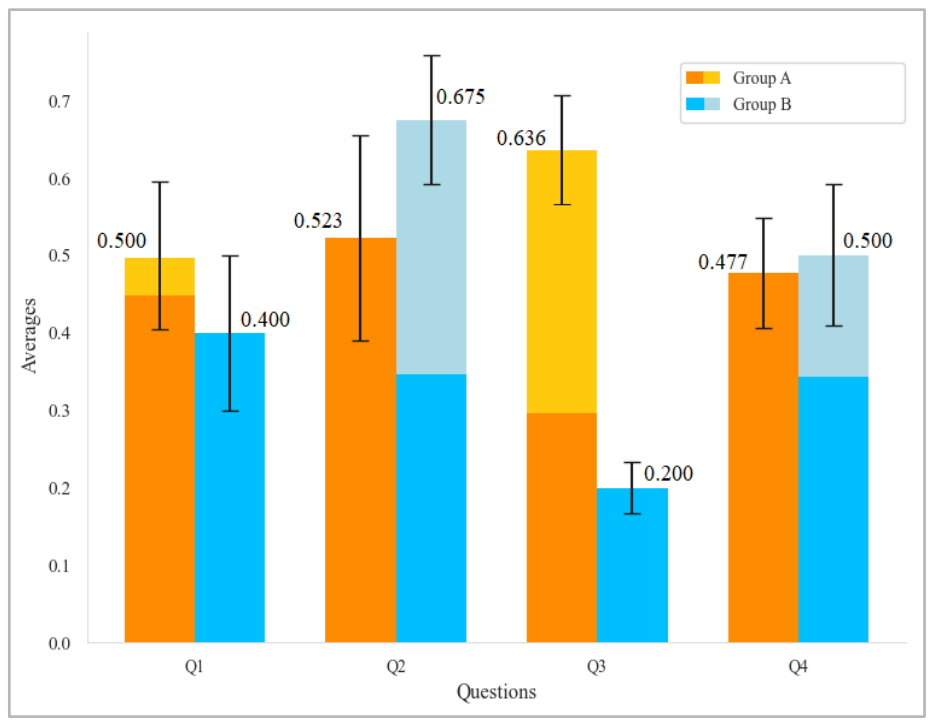

**Figure 4** The average scores for each question and the change in score when feedback was received. Dark bars show the scores of the initial responses; light bars show the average score increases after receiving feedback; the error bars represent the standard error for the total points, and the numerical values of these data are presented in Table 4.

To conduct the statistical analysis, first, the total scores of the groups on all four questions (Q1+Q2+Q3+Q4) were compared to check for any carry-over effect. As the total scores for both groups showed a normal distribution, an independent sample *t*-test was used to compare the scores. The *p*-value was determined as 0.265 in this test. Therefore, the difference between the total scores of the two groups was not statistically significant ($p>.05$). This indicates that there was no carry-over effect.

After ensuring that no carry-over effect occurred, the *D*-scores (Q1+Q3-Q2-Q4) for both groups were calculated with Equation 6. For the comparison of the *D*-scores between the two groups, independent sample *t*-test was used. The results obtained are presented in Table 5.

**Table 5** Independent sample *t*-test results used to analyse the difference between the groups' *D*-scores, which represent the difference between their response scores with and without feedback

| Group | N | $\bar{X}$ | SD | t | p | Cohen's d |
|---|---|---|---|---|---|---|
| A | 11 | 0.136 | 0.595 | 2.721 | 0.014* | 1.189 |
| B | 10 | -0.575 | 0.602 | | | |

*\* indicates a significant ($p<.05$) difference*

As Table 5 shows, the difference between the *D*-scores for the two groups is statistically significant with a large effect ($t= 2.721$, $p = 0.014$, $d = 1.189$). The *D*-scores of Group A are higher than those of Group B; therefore, the significant difference is in favour of the questions with feedback.

To conduct a more detailed analysis of whether the observed differences in scores were due to feedback, *D*-scores were calculated for both groups' scores without feedback. In this process, the scores from the students' initial responses to the questions were considered, while the scores obtained after receiving feedback were ignored. The new *D*-scores were then analysed with the independent sample *t*-test, which resulted in a *p*-value of 0.700. Therefore, when the effect of feedback is disregarded, the *D*-scores of the groups do not show a statistically significant difference.

The average *D*-scores of the two groups without and with feedback are presented in Figure 5.

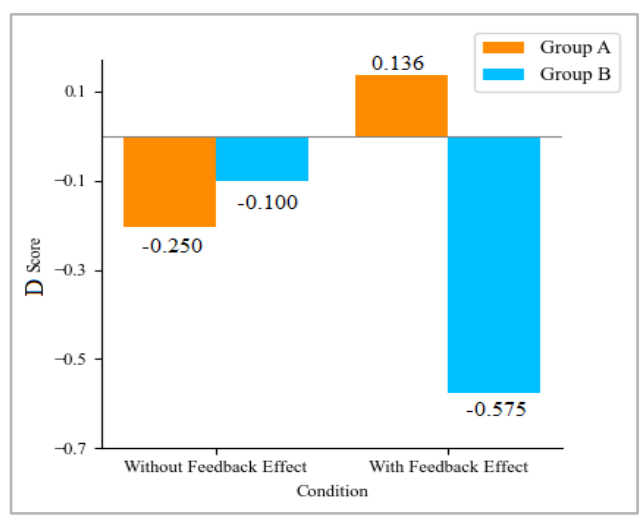

**Figure 5** *D*-score averages of the groups. The bars on the left are the *D*-scores calculated using only the scores students received from their initial answers, without feedback. The bars on the right are the *D*-scores calculated based on the total scores obtained with feedback.

As Figure 5 shows, the groups' *D*-scores are quite close to each other when the feedback effect is omitted and quite divergent when the feedback effect is included. When the effect of feedback is incorporated into the scores, the average score for Group A is 0.136, while the average score for Group B is -0.575. Group B's lower scores compared to Group A indicate that the scores for the questions with feedback are higher than those for the questions without feedback.

## 5.2 Results of cognitive learning process related to students' normalized total fixation duration during ChatGPT interactions

To achieve a more detailed analysis of the impact of ChatGPT feedback, each ChatGPT-supported question's initial score and score after receiving feedback were compared. The results of the paired sample *t*-test for this comparison are presented in Table 6.

**Table 6** Paired sample *t*-test results comparing scores for each question with and without ChatGPT feedback

|  |  | N | $\bar{X}$ | SD | t | p | Cohen's d |
|---|---|---|---|---|---|---|---|
| **Q1** | W | 11 | 0.454 | 0.395 | -1.000 | 0.170 | - |
|  | F | 11 | 0.500 | 0.362 |  |  |  |
| **Q2** | W | 10 | 0.350 | 0.428 | -3.545 | 0.003* | 0.913 |
|  | F | 10 | 0,675 | 0.265 |  |  |  |
| **Q3** | W | 11 | 0.295 | 0.101 | -4.892 | <0.001* | 1.892 |
|  | F | 11 | 0.636 | 0.234 |  |  |  |
| **Q4** | W | 10 | 0.350 | 0.316 | -1.964 | 0.040* | 0.495 |
|  | F | 10 | 0.500 | 0.289 |  |  |  |

*\* indicates a significant (p<.05) difference; (Q: Question, W: Without feedback, F: With feedback)*

Examining Table 6 reveals significant differences in favour of the post-feedback condition for Questions 2, 3 and 4. To further explore the impact of ChatGPT feedback on the response process that contributed to these differences, an eye gaze analysis was conducted. This analysis examined students' visual attention before and after receiving feedback, specifically during the first feedback session, as the goal was to directly compare the data without feedback and with feedback. The second and third feedback sessions are likely to carry over the effects of the first feedback session.

In this section, the results concerning the normalized total fixation durations during the response processes for Q1, Q2, Q3 and Q4, following the research design flow outlined in Figure 3, will be discussed in detail. The main aim of the second research question was to analyse the effect of ChatGPT feedback on visual attention. To achieve this objective, the students' visual attention to the physical AOIs of the physical experiment set and the virtual AOIs of the AR environment were compared. The effect of feedback on the differences in attention paid to these two types of AOIs was also analysed. To achieve realistic results for this research question, only the eye gaze data of the group members who used the ChatGPT function were used in the analysis.

For the first question, five groups received feedback (see Table 3). Figure 6 illustrates the distribution of the normalized total fixation duration for these groups before and after receiving feedback.

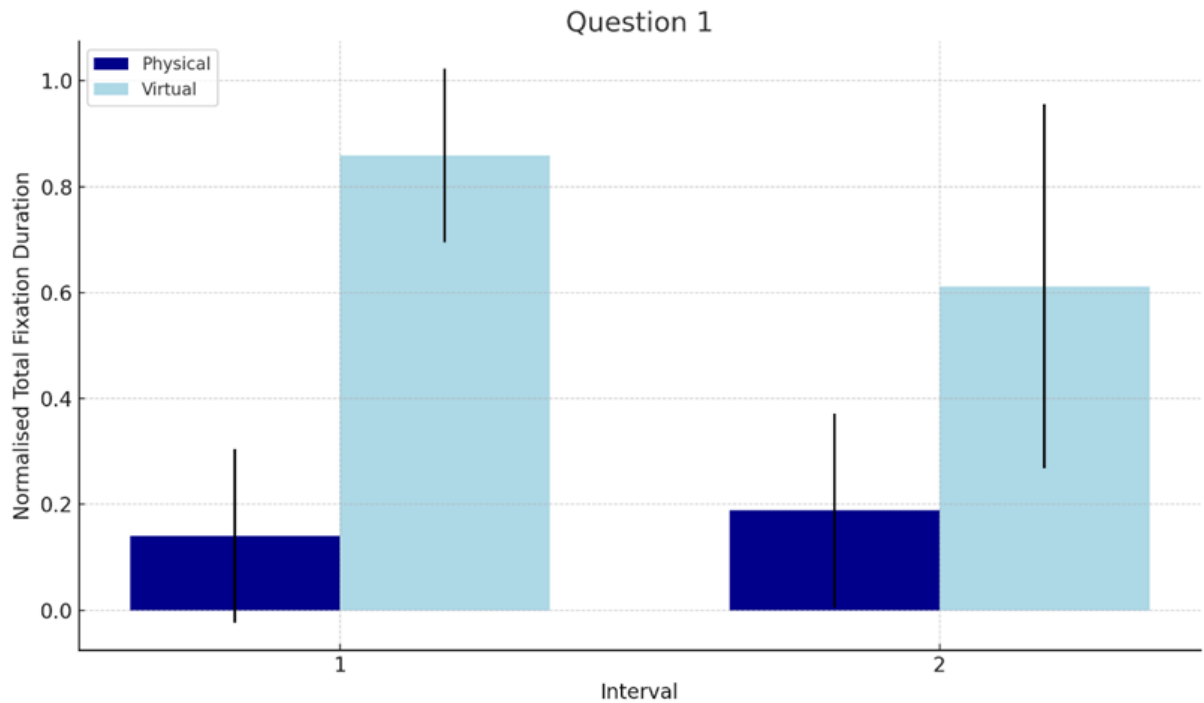

**Figure 6** Plots of the average normalized total fixation duration when answering the first question. Interval 1 represents the pre-feedback period, while Interval 2 represents the post-feedback period. The bars indicate the group average normalized total fixation duration and the standard deviation.

A paired samples *t*-test was performed to evaluate whether a difference exists between the normalized total fixation durations on physical and virtual components. The results are presented in Table 7.

**Table 7** Paired sample *t*-test results comparing the normalized total fixation durations on physical and virtual AOIs while answering the first question.

|  | AOI | N | $\bar{X}$ | SD | *t* | *p* | Cohen's d |
|---|---|---|---|---|---|---|---|
| **Interval 1** | Physical | 5 | 0.141 | 0.184 | -4.372 | 0.012* | 3.902 |
|  | Virtual | 5 | 0.859 | 0.184 |  |  |  |
| **Interval 2** | Physical | 5 | 0.188 | 0.204 | -2.237 | 0.089* | 1.379 |
|  | Virtual | 5 | 0.612 | 0.384 |  |  |  |

*\* indicates a significant (*p*<.10) difference*

The results indicate that the normalized total fixation duration on virtual components is significantly higher than that on physical components during both intervals.

For the second question, six groups received ChatGPT feedback. Figure 7 shows the distribution of their normalized total fixation durations on physical and virtual components during Interval 1 and Interval 2.

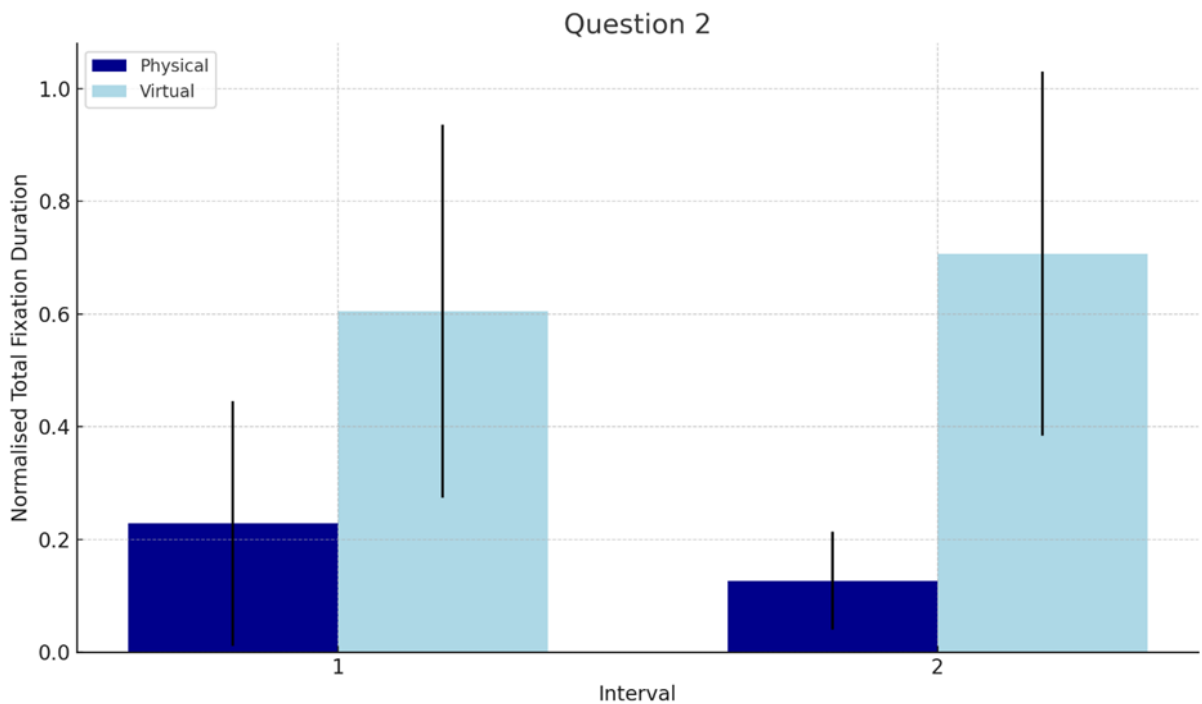

**Figure 7** Plots of the average normalized total fixation duration when answering Q2. Interval 1 represents the pre-feedback period, while Interval 2 represents the post-feedback period. The bars indicate the group average normalized total fixation duration and the standard deviation.

A paired samples *t*-test was performed to evaluate whether a difference exists between the normalized total fixation durations on physical and virtual components during the two intervals. The results are presented in Table 8.

**Table 8** Paired sample *t*-test results comparing the normalized total fixation durations on physical and virtual AOIs while answering the second question.

|  | AOI | N | $\bar{X}$ | SD | *t* | *p* | Cohen's *d* |
|---|---|---|---|---|---|---|---|
| **Interval 1** | Physical | 6 | 0.228 | 0.238 | -2.014 | 0.099* | 1.228 |
|  | Virtual | 6 | 0.605 | 0.363 |  |  |  |
| **Interval 2** | Physical | 6 | 0.127 | 0.095 | -4.460 | 0.007* | 2.238 |
|  | Virtual | 6 | 0.707 | 0.354 |  |  |  |

*\* indicates a significant (p<.10) difference*

The results indicate that the normalized total fixation duration on virtual components is significantly higher than that on physical components during both intervals.

For the third question, one group was excluded from the analysis due to technical problems during the intervention. Eye gaze analyses for the third question were performed specifically for the six groups who received feedback. Figure 8 illustrates the distribution of their normalized total fixation durations on physical and virtual components during Interval 1 and Interval 2.

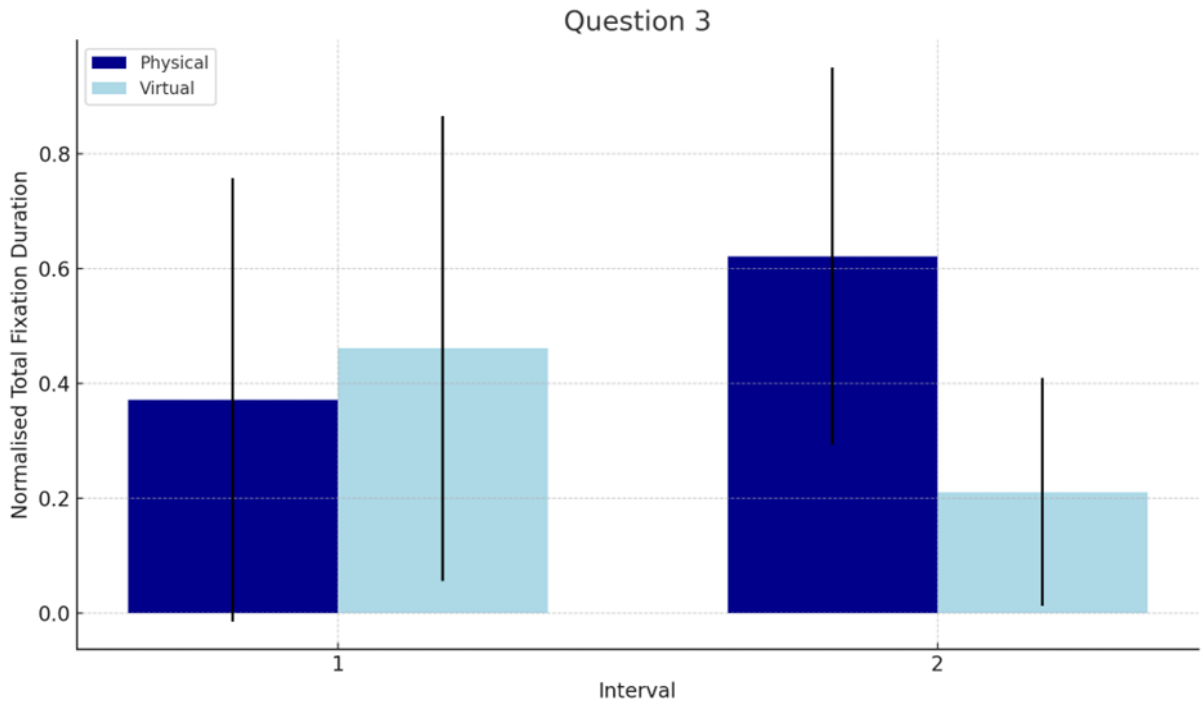

**Figure 8** Plots of the average normalized total fixation duration when answering the third question. Interval 1 represents the pre-feedback period, while Interval 2 represents the post-feedback period. The bars indicate the group average normalized total fixation durations and the standard deviation.

A paired samples *t*-test was performed to evaluate whether a difference exists between the normalized total fixation durations on physical and virtual components during the two intervals. Table 9 presents the results.

**Table 9** The paired sample *t*-test results comparing the normalized total fixation durations on the physical and virtual AOIs while answering the third question.

|  | AOI | N | $\bar{X}$ | SD | t | p | Cohen's d |
|---|---|---|---|---|---|---|---|
| **Interval 1** | Physical | 6 | 0.372 | 0.423 | -0.285 | 0.785 | - |
|  | Virtual | 6 | 0.462 | 0.444 |  |  |  |
| **Interval 2** | Physical | 6 | 0.622 | 0.360 | 2.233 | 0.068* | 1.379 |
|  | Virtual | 6 | 0.211 | 0.218 |  |  |  |

*\* indicates a significant (*p*<.10) difference*

The *t*-test results indicated that no significant differences exist between the normalized total fixation duration on virtual and physical components during Interval 1. However, after ChatGPT feedback, during Interval 2, the normalized total fixation duration on physical components is significantly higher than that on virtual components.

A total of three groups received feedback on Q4. Figure 9 illustrates the distribution of the groups' normalized total fixation durations on physical and virtual components during Interval 1 and Interval 2.

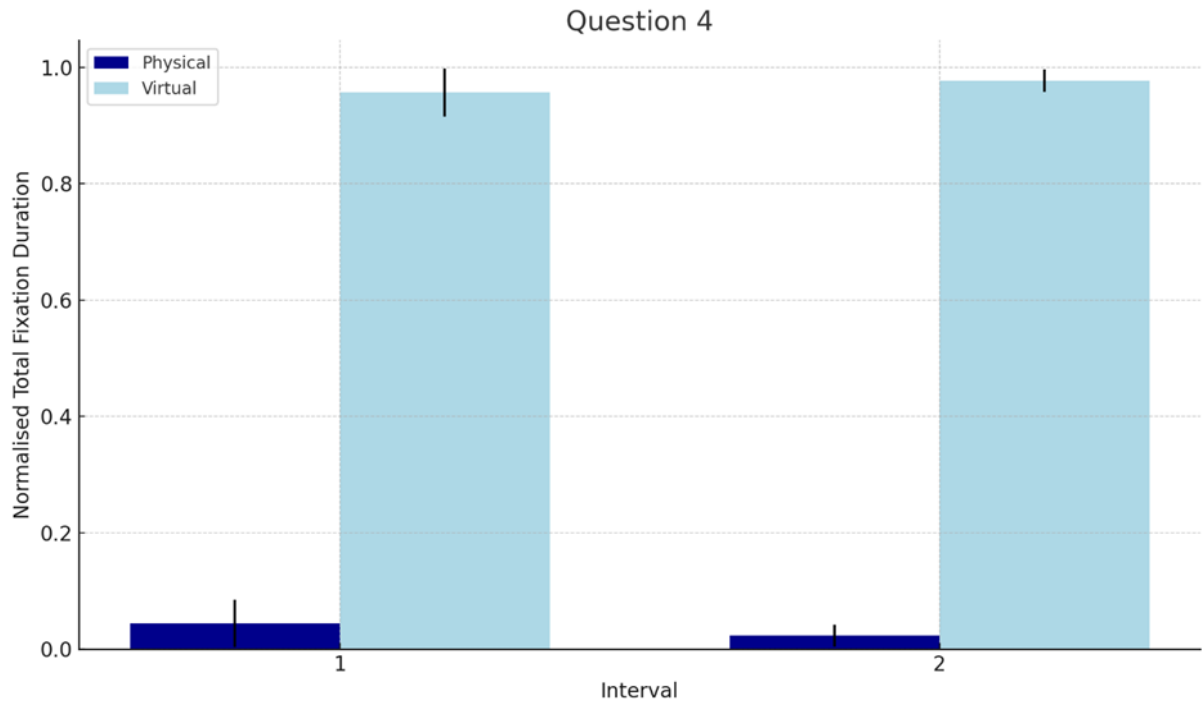

**Figure 9** Plots of the average normalized total fixation duration while answering the fourth question. Interval 1 represents the pre-feedback period, while Interval 2 represents the post-feedback period. The bars indicate the group average normalized total fixation durations and the standard deviation.

A paired samples *t*-test was performed to evaluate whether a difference exists between the normalized total fixation durations on physical and virtual components during the two intervals. Table 10 presents the results, which indicate a significant difference in the distribution of gaze during both intervals. The virtual elements received significantly more visual attention than the physical elements.

**Table 10** Paired sample *t*-test results comparing the normalized total fixation durations on the physical and virtual AOIs while answering the fourth question.

|  | AOI | N | $\bar{X}$ | SD | *t* | *p* | Cohen's *d* |
|---|---|---|---|---|---|---|---|
| **Interval 1** | Physical | 3 | 0.044 | 0.050 | -15.813 | 0.004* | 18.240 |
|  | Virtual | 3 | 0.956 | 0.050 |  |  |  |
| **Interval 2** | Physical | 3 | 0.023 | 0.023 | -35.774 | 0.001* | 41.478 |
|  | Virtual | 3 | 0.977 | 0.023 |  |  |  |

*\* indicates a significant (p<.10) difference*

The results indicated that the normalized total fixation duration on virtual components is significantly higher than on the physical components for both intervals.

## 6 Discussion

### 6.1 Discussion of the learning outcomes related to ChatGPT feedback

To answer the first research question, the differences in scores on questions with and without feedback were compared. The results showed that ChatGPT feedback had a significant positive impact on students' scores. The average scores on questions with feedback are consistently higher than those on questions without feedback.

Feedback increases students' scores on each question. The biggest difference between the feedback group and the without-feedback group was observed on Question 3, which asked about the principle Eve used to obtain information. The second highest difference in scores occurred on Question 2, which asks about the features an encryption key should have. The difference in scores was lower on Question 1, which asks why two bases are used. On Question 4, which asks about the method Alice and Bob follow to determine whether an intercepting agent is present, the average scores of the two groups exhibit almost no difference (see Table 4 and Figure 4).

Most groups answered the first and fourth questions without receiving feedback (see Table 3), and ChatGPT's feedback had a minimal effect on their scores compared to the second and third questions (see Figure 4). During the application's programming phase, the responses to the first and fourth questions in the prompts sent to ChatGPT were provided in a single paragraph without clearly distinguishing the sub-sections. This likely led ChatGPT to interpret incomplete answers as correct. Consequently, students moved on without receiving any feedback. In contrast, for the second and third questions, the sub-parts of the answers were clearly separated by item symbols, resulting in more students receiving feedback and, subsequently, achieving higher scores. This situation highlights the importance of clearly distinguishing the content within responses provided in prompts used during the programming phase to ensure effective use of ChatGPT.

One of the key points to highlight in this study is that ChatGPT was not used to provide information but rather as a tool to guide students towards discovering information. The ChatGPT feedback did not contain any direct information related to the correct answers. To ensure this, the prompts sent to ChatGPT were meticulously prepared to ensure that it avoided offering any answer-related information. Instead, the feedback was designed to encourage students to think more deeply and foster group discussions about missing points. Students used ChatGPT as a virtual group partner, and the feedback they received drew them to focus more on the points they had probably missed in their pre-feedback responses. During the post-feedback intervals, they engaged in more intense and small-scale discussions. As noted in the literature [24], these small discussions increased student success. In addition, as the literature suggests, students improved their learning outcomes in this study through ChatGPT-supported small discussions.

ChatGPT's ability to provide instant feedback has been the subject of numerous studies [53-56]. In parallel with these, the findings of this study demonstrate that ChatGPT feedback improved students' responses to the four questions asked throughout the quantum cryptography analogy experiment.

**6.2    Discussion of the learning process related to the normalized total fixation duration**

On the first, second and fourth questions, visual attention to virtual components was significantly higher during both intervals. These questions were more theoretical and conceptual in nature. Research suggests that virtual experimental processes more effectively support conceptual understanding than hands-on processes [85-88]. Therefore, in an experimental process involving both physical and virtual components, focusing more on virtual representations while answering conceptual questions may facilitate providing correct answers. The observation that students paid significantly greater attention to virtual representations when answering these concept-based questions, resulting in improved scores, supports this conclusion. In addition, ChatGPT feedback played an important role in keeping students' attention on the virtual representations, ensuring that they remained focused after receiving feedback.

In the third question, no statistical differences were observed during the first interval, but in the second interval, after receiving feedback, a significant increase in visual attention to the physical components was observed. When students initially encountered the question, they showed similar levels of attention to both types of AOIs as they responded. However, after receiving feedback, they shifted their attention to the physical components. The third question asked about Eve's behaviour during the experimental process, and the knowledge assessed by this question was technical and related to the experiment. Research shows that hands-on activities are significantly more effective for developing technical knowledge about experimental tools [85-87, 89]. Therefore, increased attention to the physical experiment setup is expected to lead to more accurate answers to technical questions. Consequently, the shift in visual focus after receiving ChatGPT feedback can be seen as supporting correct answers to this question. The change in attention between intervals demonstrates how effective ChatGPT feedback was in guiding students' attention to improve the accuracy of their responses.

## 6.4    Limitations

This study was limited to 38 students. It was conducted entirely through their voluntary participation and was not associated with any specific course curriculum. While this approach ensured that the findings were based on genuine, unbiased experiences, it also means that the sample may not fully represent the broader student population. As participants engaged in the study out of personal interest and curiosity, selection bias may have occurred.

The conversations conducted with ChatGPT were confined to the state of GPT-4 between March 6$^{th}$ and May 23$^{rd}$, 2024. Therefore, the accuracy of the feedback provided by ChatGPT is limited by the quantum cryptography and BB84 protocol knowledge of the version available during this period.

None of the participating students reported prior experience using HL2. Many students initially struggled to use the buttons within the application. However, they quickly adapted and continued to use it without issue during the subsequent sections. In addition to their lack of experience, most students mentioned that towards the end of the experimental process, the weight of the HL2 glasses and the heat from the battery started to cause some discomfort in their heads.

## 7    Conclusions

The first research question of this study considered the impact of ChatGPT feedback on student responses. The findings demonstrated a significant higher total score if the students received ChatGPT feedback. This suggests that ChatGPT feedback improved students' learning outcomes and the quality of their responses. These results support the integration of AI-driven feedback systems in educational AR settings by demonstrating that such tools can significantly improve learning outcomes.

The second research question of this study investigated the effects of ChatGPT feedback on students' gaze behaviour. ChatGPT feedback tailored to the students' responses had varying effects on their visual attention. This is an important finding in the context of personalized learning environments. For some questions, ChatGPT feedback effectively maintained students' existing visual focus. However, for one question, ChatGPT feedback shifted students' visual attention, resulting in a significant difference in attention to physical and virtual AOIs. Considering the question content and the expected gaze behaviours, ChatGPT feedback produced the desired effect on all four questions. It kept students' attention on virtual components for conceptual questions, while for the question related to the experimental process, it shifted their focus towards the physical AOIs. This indicates that ChatGPT can direct students' visual attention via personalized feedback during the question-answering process.

This study was a pilot study to determine whether and how ChatGPT feedback influences visual attention within an AR environment. Such an effect was identified, and in future research, this effect will be analysed more comprehensively across more participants. This will allow for a more detailed examination of how ChatGPT feedback affects eye-tracking data in AR environments.

Educational processes must closely follow and effectively utilize emerging technologies to cultivate individuals who will contribute to the further development of these technologies. Although this field has a limited number of studies, they indicate important effects due to mixed reality and artificial intelligence integration [90-93]. The quantitative results obtained in this study similarly demonstrate that the use of LLMs in AR environments improves student performance outcomes. Furthermore, findings related to student gaze data indicate that LLM feedback plays a crucial role in personalized learning and guiding students' focus towards more important experimental components for answering questions. Future research focusing on various uses of LLMs in AR/VR environments and their effects on different dependent variables would deepen the literature. Furthermore, eye gaze analyses in these environments are needed to conduct more comprehensive analyses and draw inferences regarding the effects of LLMs on student gaze patterns.

## 8　Controlled Variables

The system usability scale (SUS) was administered at the end of the experiment to assess students' perceptions of the usability of the experimental tools. An analysis of the results revealed an average score of 76.316 (SD: 10.295). This finding indicates that students rated the usability of the application as 'Excellent' [94], suggesting that there was no usability-related limitation of the study.